\documentclass[12pt]{article}
\usepackage{graphicx}
\usepackage{scicite}
\usepackage{times}
\usepackage{txfonts}
\usepackage{color}
\usepackage{tikz}
\usepackage{hyperref}
\usepackage{aas_macros}
\newenvironment{sciabstract}{%
\begin{quote} \bf}
{\end{quote}}

\newcounter{lastnote}

\title{A low upper limit on the subsurface rise speed of solar active regions}
\author
{Aaron C. Birch$^{1,\ast}$, Hannah Schunker$^{1}$, Douglas C. Braun$^{2}$, Robert Cameron$^{1}$, \\
Laurent Gizon$^{1,3,4,5}$, Bj\"orn L\"optien$^{1,3}$,  Matthias Rempel$^{6}$\\
\\
\normalsize{$^{1}$Max-Planck-Institut F\"ur Sonnensytemforschung, Justus-von-Liebig-Weg 3, 37077 G\"ottingen, Germany}\\
\normalsize{$^{2}$NorthWest Research Associates, 3380 Mitchell Lane, Boulder, CO 80301, USA}\\ 
\normalsize{$^{3}$Institut f\"ur Astrophysik, Georg-August-Universit\"at G\"ottingen, 37077, G\"ottingen, Germany} \\
\normalsize{$^{4}$Center for Space Science, New York University Abu Dhabi, PO Box 129188, Abu Dhabi, UAE } \\
\normalsize{$^{5}$National Astronomical Observatory of Japan, Mitaka, Tokyo 181-8588, Japan } \\
\normalsize{$^{6}$National Center for Atmospheric Research, High Altitude Observatory, 3080 Center Green Drive,}\\
\normalsize{Boulder, CO 80301, USA}\\
\normalsize{$^\ast$Corresponding author; E-mail:  birch@mps.mpg.de}
}
\date{}

\begin{document}

\baselineskip24pt

\maketitle

\begin{sciabstract}
Magnetic field emerges at the surface of the Sun as sunspots and active regions. This process generates a poloidal magnetic field from a rising toroidal flux tube; it is a crucial but poorly understood aspect of the solar dynamo. The emergence of magnetic field is also important because it is a key driver of solar activity. We show that measurements of horizontal flows at the solar surface around emerging active regions, in combination with numerical simulations of solar magnetoconvection, can constrain the subsurface rise speed of emerging magnetic flux. The observed flows imply that the rise speed of the magnetic field is no larger than 150~m/s at a depth of 20~Mm, that is, well below the prediction of the (standard) thin flux tube model but in the range expected for convective velocities at this depth. We conclude that convective flows control the dynamics of rising flux tubes in the upper layers of the Sun and cannot be neglected in models of flux emergence.
\end{sciabstract}

\section*{Introduction}

Solar active regions are thought to be the surface manifestation of magnetic flux tubes emerging from the solar interior  \cite{Parker1955}. These flux tubes are thought to be formed deep in the Sun, in the stably stratified layer just beneath the convection zone  \cite{Spiegel1980}. This is the current prevailing picture and has been used to explain (i) the latitudes at which bipolar active regions emerge  \cite{Zwaan1978} and (ii) Joy's law: the tendency for the leading polarity to be closer to the equator than the trailing polarity \cite{Fan2009, Weber2013}. Alternative views are that active regions are the consequence of a dynamo operating in the shallow solar interior \cite{Brandenburg2005} or the bulk of the convection zone, for example, in the work of Nelson et al. \cite{Nelson2013}.

Understanding the physics of magnetic flux emergence is crucial to understanding the conversion of toroidal to poloidal magnetic field through the tilting of active regions (Joy's law). In addition, magnetic flux emergence plays a central role in driving solar activity, a topic of very broad interest in solar physics \cite{Cheung2014}. In an even broader context, understanding flux emergence on the Sun may play an important role in understanding stellar activity in general \cite{Baliunas1985}.

Thin flux tube models predict that magnetic flux concentrations originating from the bottom of the convection zone reach upward speeds of about 500 m/s at 20 Mm below the surface and then accelerate rapidly as the flux tube approaches the surface  \cite{Fan2009}. Three-dimensional (3D) anelastic simulations have also been used to model the rise of magnetic flux concentrations through the convection zone \cite{Fan2008,Jouve2009,Jouve2013}. These simulations are carried out in a computational domain with a top boundary at 20 to 30 Mm below the solar surface, and thus, it is not possible to make direct observational contact with these simulations. Within 20 Mm of the surface, the thin flux tube approximation is not justified because the tube radius is no longer small compared to the scale height of the solar stratification and the anelastic approximation is not justified due to compressibility, and realistic numerical simulations of magnetoconvection are required instead \cite{Cheung2010, Stein2012, Rempel2014}.

The upper convection zone can be probed by helioseismology -- the study of solar oscillations to learn about the solar interior [see the review by Gizon et al. \cite{Gizon2010}]. Helioseismology has been used in the past to search for signatures of the magnetic flux concentrations below the surface before active region emergence \cite{Kosovichev2000,Zharkov2008,Ilonidis2011}. This approach is promising but challenging \cite{Birch2013,Braun2012,Ilonidis2012}.

Here, we take a new approach and use observations of surface flows together with numerical simulations of solar magnetoconvection to constrain physical models of the rising magnetic flux concentrations that form active regions. This approach has not been used before; it has only recently become feasible as realistic numerical simulations have become possible  \cite{Rempel2014}.

\section*{Results}

The Helioseismic and Magnetic Imager (HMI)  \cite{Scherrer2012, Schou2012} on the Solar Dynamics Observatory (SDO), launched in 2010, has provided observations of the full visible solar disc with almost complete temporal coverage. From the active regions observed by HMI/SDO to emerge on the visible disc in the time period from April 2010 to November 2012, we selected 70 active regions without strong pre-existing magnetic flux near the emergence location. All of these active regions have a NOAA (National Oceanic and Atmospheric Administration) active region number and emerge into the quiet Sun. These regions range from small active regions that barely form a sunspot to very large active regions with a number of sunspots (total unsigned line-of- sight flux varies from $10 \times 10^{20}$ to $400 \times 10^{20}$~Mx). For each of these emerging active regions, we also identified a partner quiet-Sun control region with the same disc position but at a different time. These regions serve as a control sample to ensure that any observed flow signatures are due to flux emergence rather than systematic effects associated with disc position.

\subsection*{Measurements of surface flows}

We used both helioseismology and local correlation tracking (LCT) of granulation to measure the horizontal surface flows associated with each of these emerging active regions. Using two independent methods allows us to validate the horizontal flow measurements on the relevant spatial scales of several megameters and larger. For the helioseismology, the input data are the 45-s cadence Doppler images \cite{Couvidat2012}, and we used helioseismic holography \cite{Lindsey2000} to infer near-surface horizontal flows (we used helioseismic measurements that are sensitive to flows in the top few megameters of the convection zone). The LCT was done using the Fourier LCT (FLCT) code  \cite{Fisher2008, Welsch2004} to track the granulation seen in the intensity images. In both cases, we obtained flow maps with a temporal cadence of about 6 hours for each of the active regions. Comparison of the helioseismology and LCT results confirms that the two methods are measuring the same horizontal flows at the surface. In addition, we used the line-of-sight magnetograms from HMI to follow the magnetic evolution of each region, also with 6-hour time resolution.

\subsection*{Simulations of rising magnetic flux concentrations}
\begin{figure}
\includegraphics[width=0.8\linewidth]{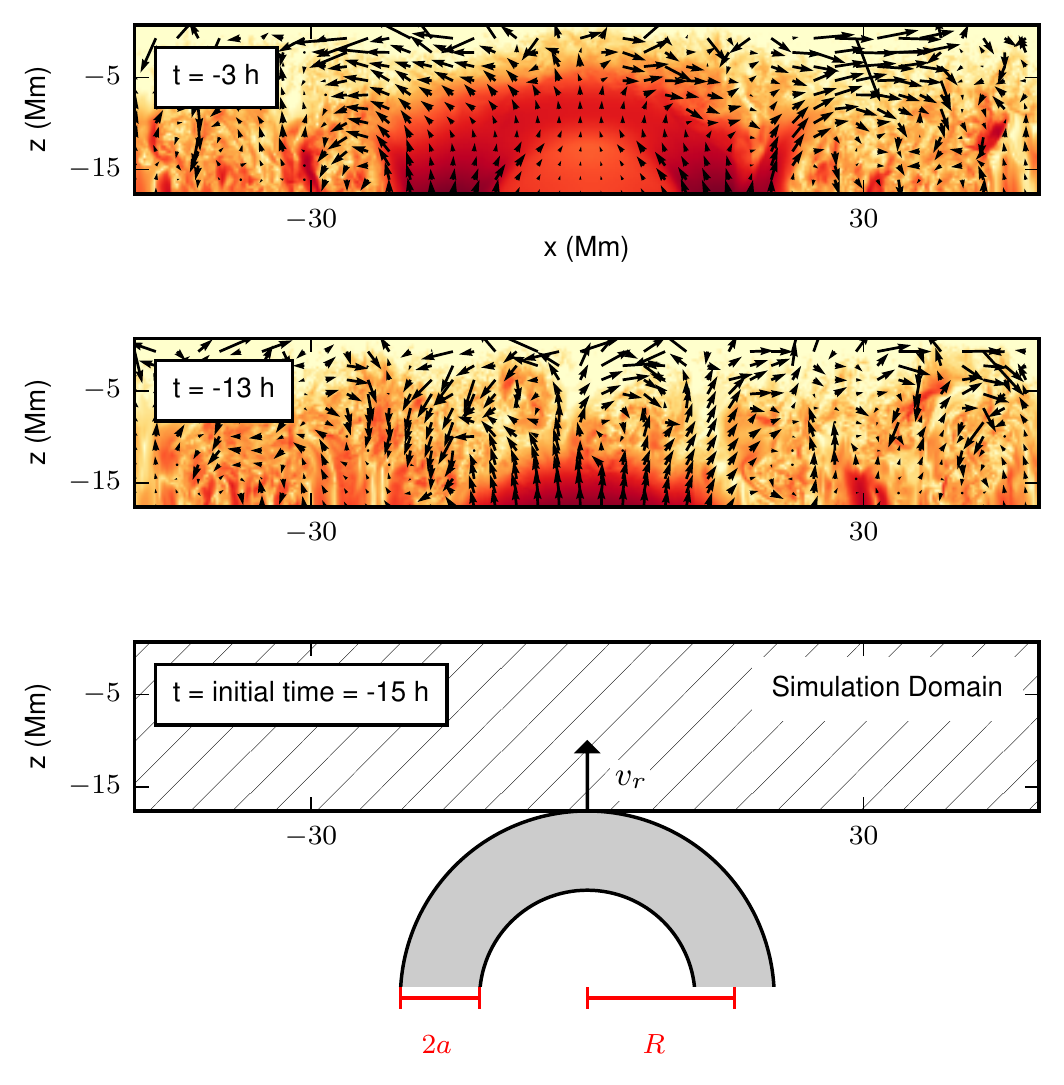}
\caption{Diagram of the setup for the simulation (bottom panel) and vertical slices through the simulation with a rise speed of 500~m/s at 13 and 3~hours before the emergence time (middle and top panels). In the bottom panel, the magnetic flux tube used to generate the bottom boundary condition is shown in gray. The major radius of the flux tube is given by $R$, the minor radius is given by $a$, and an untwisted magnetic field is oriented along the tube. Within the minor radius, the magnetic field has a Gaussian dependence of the form $\exp{[ - 2 x^2/a^2 ]}$, where $x$ is the distance from the center of the tube; the magnetic field is zero outside the tube. As the simulation progresses, the flux tube moves upward with rise speed $v_r$. In the top two panels, the log of the magnetic field strength is color-coded (red is the strongest field, and light yellow is the weakest), and the arrows show the flows in the plane of the vertical slice (the largest arrows represent about 3~km/s). Upward and horizontal diverging flows are apparent during the emergence process in this case.  \label{fig.fig1}}
\end{figure}

To constrain the physics of the magnetic flux emergence process, we carried out a series of comprehensive radiative magnetoconvection simulations of magnetic flux concentrations rising through the top 20 Mm of the solar convection zone. We used the MURaM code  \cite{Vogler2005} with only minor modifications to the setup of Rempel and Cheung \cite{Rempel2014}. The domain size was $98 \times 98 \times 18$~Mm$^3$, and each run was carried out for 100 hours of solar time. In all cases, we emerged a half torus of magnetic flux as detailed in the work of Rempel and Cheung \cite{Rempel2014}and shown schematically in Fig.~\ref{fig.fig1}. For the cases shown here, we used a major radius $R = 16$~Mm, a minor radius $a = 6.1$~Mm, and a field strength of 20~kG on the axis of the torus. The magnetic field is untwisted and has a Gaussian profile with distance from the axis of the torus. The magnetic field strength is $20$~kG~$\times \exp{[-2]} \approx 2.7$~kG at the boundary of the flux tube. The total flux is $10^{22}$~Mx, and the average field strength is about 10~kG. These parameters for the initial magnetic configuration are plausible and are consistent with typical values of the observed magnetic flux of active regions and magnetic field strengths from thin flux tube calculations. We vary the imposed rise speed of the magnetic flux at the bottom boundary from 70 to 500~m/s in a series of simulations. For comparison, the average upflow convective velocity at the bottom of the domain is about 140 m/s. We used the vertical velocity (at optical depth $\tau = 0.01$) from the simulations as proxy Doppler images to measure the horizontal flows in the simulations using helioseismology. This allows direct comparison with the observations.

\subsection*{Comparison of observed and simulated surface flows}
Quantitative comparison of observations and simulations requires a consistent definition of emergence time and emergence location. Following Leka et al., \cite{Leka2013} we define the emergence time as the time when the increase in magnetic flux from its quiet-Sun background level reaches 10\% of its maximum. The emergence location is computed as the centroid of the pixels at which the change in the magnetic field from 24 hours before emergence to 8 hours after emergence is more than 30\% of the maximum change. Active region emergences, in general, have a wide variety of characteristics (for example, multiple emergence episodes in time and space). We selected these simple definitions for ease of reproducibility [see the work of Leka et al.  \cite{Leka2013} for further discussion].

\begin{figure}
\includegraphics[width=\linewidth]{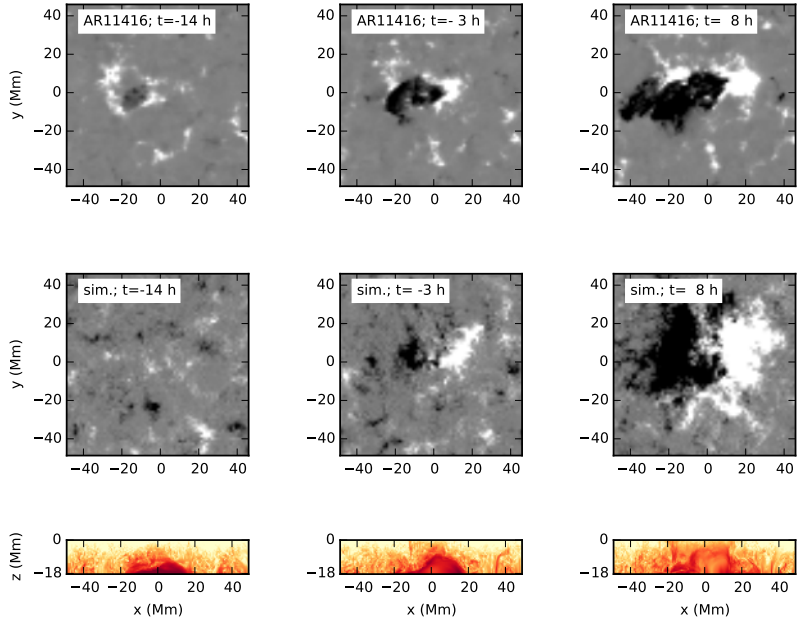}
\vspace*{-2.5in}
\caption{Simulations of the emergence of magnetic flux through the photosphere reproduce many of the features of observed emergences. The top row shows, from left to right, the time evolution of the line-of-sight surface magnetic field associated with the emergence of AR11416 as seen by SDO/HMI. The images cover the time period from 14 hours before the emergence time to 8 hours after the emergence. The middle row shows the vertical surface magnetic field of a simulation for the case with a flux tube rising at a rise speed of 140 m/s. The bottom row shows vertical slices at $y = 0$ through the magnetic field strength in the same simulation. In the first two rows, positive (negative) line-of-sight magnetic field is shown in white (black). The gray scale is saturated at 120 G. In the bottom row, the color scale is the same as in Fig.~\ref{fig.fig1}\label{fig.fig2}}
\end{figure}

We applied these definitions to compute emergence times and locations for the observed active regions as well as the simulations. Figure~\ref{fig.fig2} shows the evolution of the photospheric line-of-sight magnetic field for the observed AR11416 and the vertical magnetic field in the simulation for a flux tube with a rise speed of 140 m/s. The time evolution of the magnetic fields in the observations and in the simulation are qualitatively similar, although the observations show hints of the emerging magnetic field at 14 hours before emergence, which is not seen in the simulation.

\begin{figure}
\includegraphics[width=\linewidth]{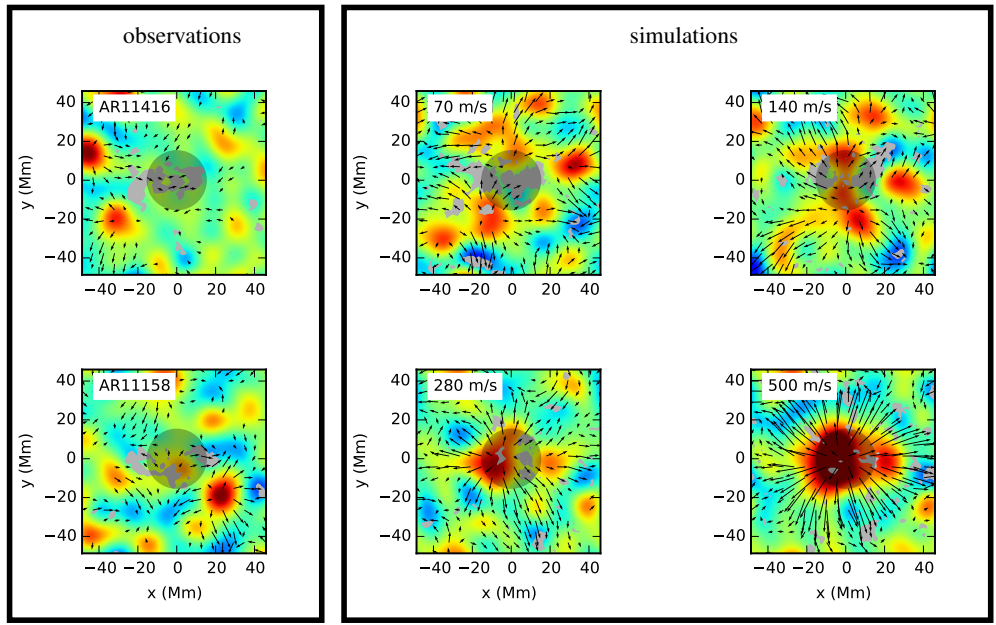}
\vspace*{-2.5in}
\caption{ At 3 hours before the emergence time,the near-surface flows inferred from HMI observations are dominated by convection, where as the simulations show a diverging flow that increases in strength with the rise speed of the flux tube at 20-Mm depth. The two panels in the left column show maps of the horizontal divergence of the flows measured from helioseismology (red for diverging flows and blue for converging flows), flow maps from helioseismology (black arrows), and line-of-sight magnetic field strength (from magnetograms, shaded gray regions for fields stronger than 60 G) for the HMI observations of AR11416 and AR11158. The four-panel group on the right shows simulations for 10-kG flux tubes with rise speeds of 70, 140, 280, and 500 m/s at 20-Mm depth. The simulations with rise speeds of 280 and 500 m/s produce strong diverging flows that are not seen in the observations. The longest arrows represent flows of 400 m/s.\label{fig.fig3}}
\end{figure}

Figure~\ref{fig.fig3} shows the helioseismically inferred near-surface flows at 3 hours before emergence for AR11416 and AR11158 (the flows from LCT are very similar; correlation coefficient of about 0.9; see fig. S1). The flow pattern is dominated by supergranulation-scale (30 Mm) flows. There is some magnetic field seen at the emergence location in both cases, but there is no clear flow pattern associated with the magnetic flux emergence. The simulations (right box in Fig. 3) show that the preemergence flow field depends on the rise speed with which the flux tube is introduced through the bottom of the simulation domain. For the case of a rise speed of 70 m/s, the flow pattern is, like the observations, dominated by the quiet-Sun convection pattern. Also, as in the observations, the magnetic field is concentrated near the emergence location. For the case of a rise speed of 140 m/s, the situation is similar, but there is a weak diverging flow away from the emergence location. For the remaining two cases, there is a strong diverging flow of several hundreds of meters per second.

\begin{figure}
\includegraphics[width=0.8\linewidth]{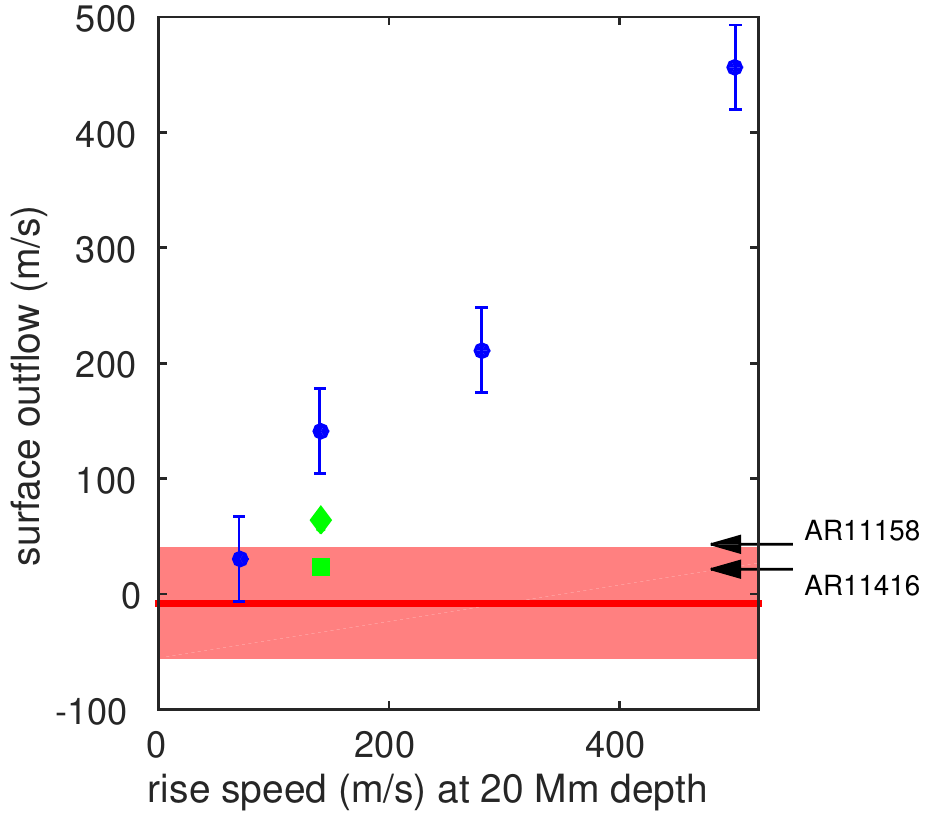}
\caption{ Comparison of the azimuthal average of the horizontal surface outflow from HMI observations with the surface outflows seen in the simulations rules out flux tubes with total flux $10^{22}$~Mx rising with speeds above 140 m/s at 20-Mm depth. The simulations (blue circles) show a radial surface outflow that increases with the rise speed of the flux tube through the bottom boundary of the simulation domain. The error bars for the simulations show upper limits on the noise in the seismology measurement procedure. The horizontal black arrows show the observations for AR11416 and AR11158 (the examples shown in Fig. 3). The red shaded region shows the one-s variations in the azimuthal average of the horizontal outflow at a distance of 15 Mm from the emergence location for the sample of 70 observed active regions. The green diamond shows the reduced field strength case (error bars not shown; the errors are the same as the other simulations), and the green square shows the reduced cross-section case. The simulations with a rise speed of 140 m/s with the tube located in the strongest upflow or strongest downflow at the bottom boundary produce diverging flows of about 90 m/s. As discussed in the text, the surface flows driven by flux emergence depend not only on the rise speed but also on the geometry and field strength of the rising flux tube.  \label{fig.fig4}}
\end{figure}

From the examples shown in Fig.~\ref{fig.fig3}, we see that the observed active region emergences are not preceded by strong diverging flows, as has been noticed before in case studies \cite{Kosovichev2009, Getling2015}. To determine whether the observed active regions shown in Fig. 3 are typical, we carried out a statistical analysis on our sample of 70 emerging active regions. Figure~\ref{fig.fig4} shows the azimuthal average of the radial outflow at a distance of 15~Mm from the emergence location at 3 hours before the emergence time. The one-$\sigma$ range allowed by observations is $-8 \pm 50$~m/s. For comparison, the one-$\sigma$ range for the quiet-Sun control regions is $-5 \pm 40$~m/s. The simulation with a flux tube rise speed of 70 m/s is consistent with observations at about the one-$\sigma$ level. The simulations with rise speeds of 280 and 500 m/s are excluded by the observations. 

For the case of a rise speed of 140 m/s, we carried out four additional simulations. In two cases, we placed the rising flux tube in the strongest upflow or downflow in the simulation domain. For the third case, we reduced the average field strength in the flux tube by a factor of 2. In the final case, we reduced the tube cross section by a factor of 2. In these last two cases (reduced field strength and reduced tube cross section), the total magnetic flux within the flux tube is half that of the other simulations. Of these, only the simulation with reduced cross section is consistent with the observations (Fig.~\ref{fig.fig4}).

\section*{Discussion}

We have shown that models with a mean initial magnetic field strength of 10 kG, an initial tube minor radius $a = 6.1$~Mm, and rise speeds at or above 140 m/s at a depth of 20 Mm produce preemergence diverging flows at the surface that are not compatible with SDO/HMI observations. The case with a rise speed of 70 m/s and also the case with a rise speed of 140 m/s, but reduced cross section, both produce flows that are weak enough to be allowed by the observations. In future work, it will be interesting to carry out a more complete exploration of the full parameter space. The upper limit on the rise speed that we have found here is about the same as the maximum speed of the convective upflows at the bottom of the simulation domain. We thus conclude that convection must play a key role in the flux emergence process.

The upper limit on the rise speed of magnetic flux concentrations obtained here ($\sim 140$~m/s) is three times smaller than the rise speed predicted by thin flux tube calculations [$\sim 500$~m/s at 20 Mm; see the work of Fan~\cite{Fan2009}]. Although thin flux tube calculations reproduce the latitudes at which active regions emerge and also their tilt angles, these models cannot address the interaction of flux tubes with convection in the near-surface layers. Future work is required to determine a scenario for the formation of magnetic flux concentrations and their journey to the solar surface that is compatible with the observations presented here.

\section*{Materials and Methods}

\subsection*{LCT to measure surface flows}
For each active region, we identified the time period of interest (381 min before the emergence time to 29 min after the emergence time) using the definition of emergence time described earlier in the paper. To account for the main effects of solar rotation, we tracked the HMI intensity images at the Carrington rotation rate. We then used a Postel projection to remap each image to a Cartesian coordinate system with a map scale of 348 km/pixel. This procedure resulted in a 3D data cube (two spatial dimensions plus time) for each emerging active region. We then applied the FLCT code \cite{Welsch2004, Fisher2008} to estimate the horizontal flows on the surface by following small-scale patterns (granulation). The FLCT requires a parameter $\sigma$; this parameter defines the size of the subregions to which the correlation tracking is applied. We chose the parameter $\sigma = 6$~pixels ($\approx2$~Mm), which is appropriate for following the granulation pattern; the choice of s is discussed in detail by L\"optien et al.~\cite{Loeptien2016}. The FLCT code provides an estimate of the surface flows with the same cadence as the input data (45 s). To obtain the surface flow maps described here, we then averaged the flows over the entire time period of interest (410 min).

\subsection*{Helioseismology of surface flows}
To prepare the input data for the helioseismology analysis, we applied the same tracking and mapping procedure as described above but used the HMI Doppler velocity images and a map scale of 1.39 Mm/pixel. We then filtered the remapped data with filter~3 from Table~1 from the work of Couvidat et al.~\cite{Couvidat2005} (this is a phase-speed filter with a central phase speed of 17.49 km/s and a width of 2.63 km/s). This filter isolates waves with a lower turning point of about 3 Mm below the photosphere. We then applied surface-focusing helioseismic holography \cite{Braun2008} to measure north-south and east-west travel time differences. These travel times are proportional to the horizontal flows in the near-surface layers. We used a conversion constant of $-5.7$~m~s$^{-2}$ to convert from travel time differences to surface flows; this empirically determined constant gives the best match between the amplitude of the surface flows measured by LCT and helioseismology.

\subsection*{MURaM simulations of emerging flux}
The MURaM radiative magnetohydrodynamic (MHD) code solves the MHD equations using a finite difference discretization that is fourth-order accurate in space and time, coupled with a short characteristic radiative transfer scheme \cite{Vogler2005}. The code uses the OPAL equation of state. Turbulent diffusivity has two components in the simulation. It is explicitly captured for the scales that are resolved and implicitly treated through a slope-limited diffusion scheme as detailed in the work of Rempel  \cite{Rempel2014b}. Because of the rather low resolution of the simulations presented here, the latter dominates in the photosphere on the scale of granulation. In the deeper parts of the domain, the contribution from the numerical diffusivity is small enough to allow for the presence of a small-scale dynamo that maintains a mixed polarity field independent from the magnetic field we emerge.

The code has an open bottom boundary condition, which allows convective flows to cross the boundary and to provide the energy flux that is required to balance radiative loss in the photosphere. To this end, the entropy is specified in upflows at the bottom boundary, whereas outflows transport out their typically lower entropy fluid. The top boundary, located about 700 km above the photosphere, is closed. The magnetic boundary condition at the bottom is such that it mirrors magnetic field components into the boundary layers [see the work of Rempel and Cheung \cite{Rempel2014} for details and implications for mixed polarity field], and at the top boundary, the magnetic field is matched to a potential field extrapolation. The boundary conditions at the bottom boundary are overwritten during flux emergence. In the flux emergence region, we imposed a specified upflow velocity $v_r$ and a magnetic field that corresponds to a semitorus advected into the domain by the velocity $v_r$ \cite{Cheung2010, Rempel2014}. We used here a “soft coupling” in which the flow and magnetic field are driven toward the imposed velocity and magnetic field on a time scale that corresponds to about $10 \; h_z/v_r$, where $h_z$ is the vertical grid spacing. The pressure at the bottom boundary was allowed to adjust to become dynamically consistent with the imposed velocity (the mean pressure at the bottom boundary is fixed, but local pressure perturbations are allowed). After the emergence was finished, we transitioned back to the open boundary described above, which did not further anchor the magnetic field and allowed for free evolution. Unlike in some of the cases \cite{Rempel2014}, there was no field-aligned flow imposed by the flux emergence boundary condition.

\section*{Acknowledgments}
{\bf Funding}: A.C.B., H.S., and L.G. acknowledge research funding by Deutsche Forschungsge- meinschaft (DFG) under grant SFB 963 “Astrophysical flow instabilities and turbulence.” D.C.B. and M.R. received support from the NASA Heliophysics Division through project awards NNH12CF23C and NNH12CF68C. The HMI data used are courtesy of NASA/SDO and the HMI science team. The data were processed at the German Data Center for SDO, funded by the German Aerospace Center (DLR). We used the workflow management system Pegasus (funded by the NSF under OCI SI2-SSI program grant \#1148515 and OCI SDCI program grant \#0722019). The National Center for Atmospheric Research is sponsored by the NSF. Computing time was provided by the NASA High-End Computing Program through the NASA Advanced Supercomputing Division at Ames Research Center under projects s1325 and s1326. {\bf Author contributions}: A.C.B., H.S., D.C.B., R.C., and L.G. jointly designed the research. H.S. and D.C.B. carried out the helioseismic flow measurements and prepared the magnetograms. B.L. carried out the LCT flow measurements. M.R. designed and ran the MURaM simulations. A.C.B. carried out the analysis of the helioseismic and LCT flow maps. The manuscript was drafted by A.C.B. with contributions from all authors. {\bf Competing interests}: The authors declare that they have no competing interests. {\bf Data and materials availability}: The SDO/HMI data used to obtain the conclusions in this paper are available at \url{http://jsoc.stanford.edu}. All data needed to evaluate the conclusions in the paper are present in the paper and/or the Supplementary Materials. Additional data related to this paper may be requested from the authors.

\makeatletter 
\setcounter{figure}{0}
\renewcommand{\thefigure}{S\@arabic\c@figure}
\makeatother

\section*{Supplementary Materials}

\begin{figure}
\includegraphics[width=0.8\linewidth]{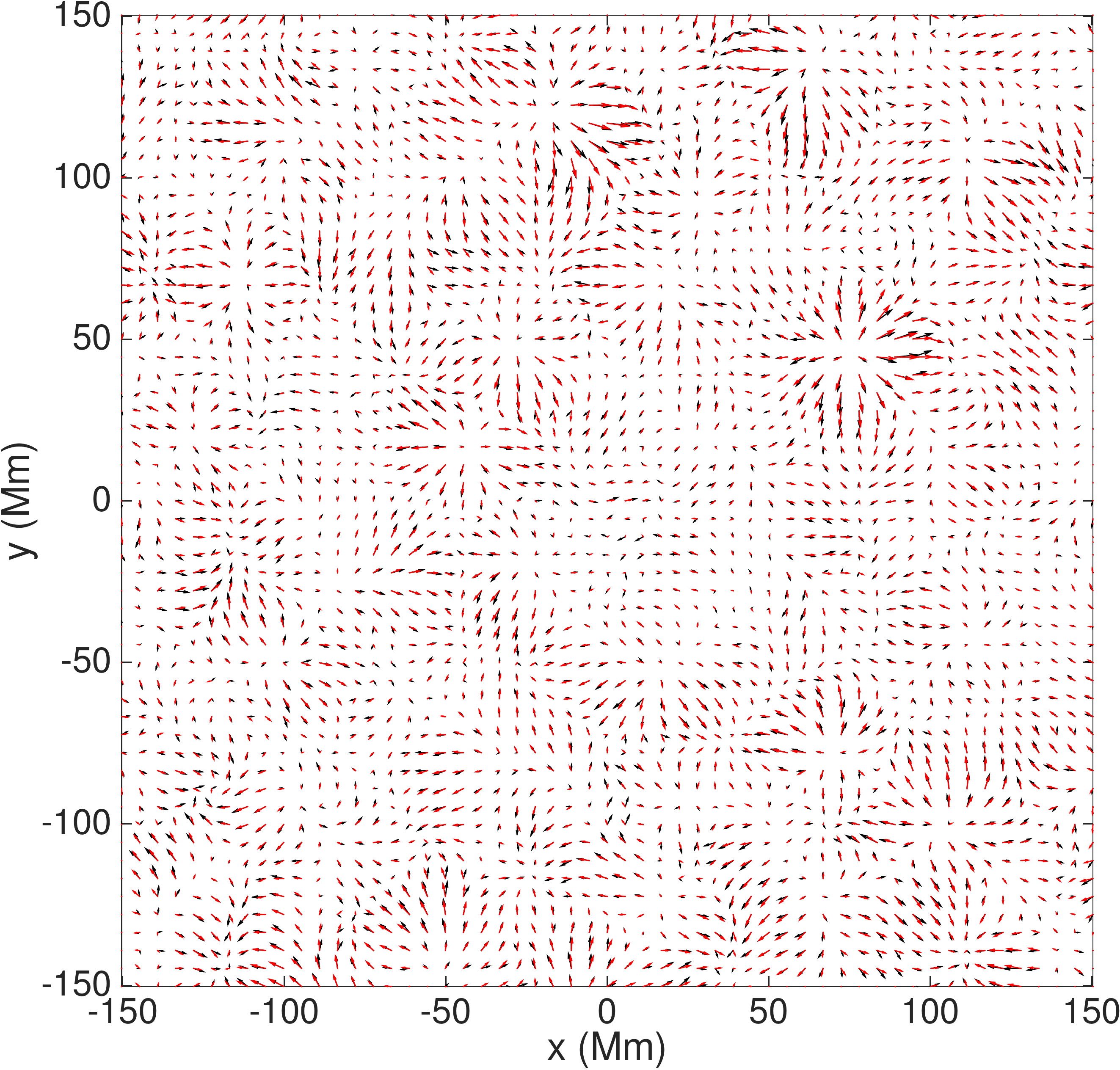}
\caption{Comparison of horizontal surface flows measured using LCT (red) and helioseismology (black) for AR11416. The field of view is much larger than in Fig.~3a. The longest arrows represent flows of about 300~m/s. The correlation coefficients between the x and y components of the flow are both about 0.93.\label{fig.sup}}
\end{figure}

The supplementary material is shown in Figure~\ref{fig.sup}.

\bibliographystyle{ScienceAdvances}
\bibliography{ms}

\end{document}